\documentclass[12pt,preprint]{emulateapj}
\usepackage{natbib}

\shorttitle{SONYC - $\rho$ Ophiuchi}
\shortauthors{Mu\v{z}i\'c et al.}

\begin{document}
\bibliographystyle{apj}


\title{Substellar Objects in Nearby Young Clusters (SONYC) V: New brown dwarfs in $\rho$ Ophiuchi \footnote{Based in part on data collected at Subaru Telescope, which is operated by the National Astronomical Observatory of Japan. Based in part on observations collected at the European Organisation for Astronomical Research in the Southern Hemisphere, Chile (program 385.C-0450).}
}


\author{Koraljka Mu\v{z}i\'c\altaffilmark{1}, Alexander Scholz\altaffilmark{2}, Vincent Geers\altaffilmark{3}, Ray Jayawardhana\altaffilmark{1,**}, Motohide Tamura\altaffilmark{4}}

\email{muzic@astro.utoronto.ca}

\altaffiltext{1}{Department of Astronomy \& Astrophysics, University of Toronto, 50 St. George Street, Toronto, ON M5S 3H4, Canada}
\altaffiltext{2}{School of Cosmic Physics, Dublin Institute for Advanced Studies, 31 Fitzwilliam Place, Dublin 2, Ireland}
\altaffiltext{3}{Institut f\"ur Astronomie, ETH, Wolfgang-Pauli-Strasse 27, 8093 Z\"urich, Switzerland}
\altaffiltext{4}{National Astronomical Observatory, Osawa 2-21-2, Mitaka, Tokyo 181, Japan}
\altaffiltext{**}{Principal Investigator of SONYC}

\begin{abstract}
SONYC -- {\it Substellar Objects in Nearby Young Clusters} -- is a survey program to investigate the frequency and properties of substellar objects with masses down to a few times that of Jupiter in nearby star-forming regions. For the $\sim$1$\,$Myr old $\rho$ Ophiuchi cluster, in our earlier paper we reported deep, wide-field optical and near-infrared imaging using Subaru, combined with 2MASS and Spitzer photometry, as well as follow-up spectroscopy confirming three likely cluster members, including a new brown dwarf with a mass close to the deuterium-burning limit. Here we present the results of extensive new spectroscopy targeting a total of $\sim$100 candidates in $\rho$ Oph, with FMOS at the Subaru Telescope and SINFONI at the ESO's Very Large Telescope. We identify 19 objects with effective temperatures at or below 3200$\,$K, 8 of which are newly identified very-low-mass probable members of $\rho$ Oph. Among these eight, six objects have $T_{\mathrm{eff}} \leq 3000$K, confirming their likely substellar nature. These six new brown dwarfs comprise one fifth of the known substellar population in $\rho$ Oph. We estimate that the number of missing substellar objects in our survey area is  $\sim 15$, down to $0.003 - 0.03 M_{\odot}$  and for $A_V = 0 - 15$. The upper limit
on the low-mass star to brown dwarf ratio in $\rho$ Oph is $5.1 \pm 1.4$, while the disk fractions are $\sim 40 \%$ and $\sim 60\%$ for stars and BDs, respectively. Both results are in line with those for other nearby star forming regions.  
\end{abstract} 

\section{Introduction}
\label{s1}

Understanding the origin of the stellar initial mass function
(IMF) is one of the major topics in astrophysics. The low-mass
end of the IMF, in particular, has been subject of numerous
observational and theoretical studies over the past decade (see \citealt{bonnell07}).
Brown dwarfs (BD; $M\lesssim 0.08\,M_{\odot}$) are an important class of objects to test the mass dependence 
in the formation and early evolution of stars and planets.  
Establishing a census of young substellar objects is a  
fundamental prerequisite to test the competing theoretical scenarios trying to explain their origin. These scenarios include turbulent fragmentation \citep{padoan&norlund04}, ejections from multiple systems \citep{bate09} or protoplanetary disks  \citep{stamatellos09}. Further issues that can be addressed by investigating young brown dwarfs are, among others, the evolution of accretion disks (e.g., the impact of photoevaporation), the formation of multiple systems \citep{bate09}, the mass dependence of planet formation 
\citep{payne&lodato07}, and the early regulation of angular momentum \citep{scholz04,scholz05}. Young BDs also provide excellent test-beds for low-gravity atmosphere models \citep[e.g.][]{kirkpatrick08}. 

Surveys in star-forming regions have revealed the existence of free-floating objects with masses
below the Deuterium-burning limit ($\sim 0.015\,M_{\odot}$; \citealt{zapateroosorio00, lucas05, jayawardhana&ivanov06, luhman08}), i.e. comparable
to the masses of giant planets. Recently, \citet{sumi11} presented evidence for the existence of a population of unbound or distant Jupiter-mass objects
in the field, that may have formed in proto-planetary disks and subsequently scattered. 
Nevertheless, current surveys in star forming regions are incomplete in the substellar regime, which is especially pronounced in the planetary-mass
domain. Some of the deepest surveys are biased by design, either because they are based on mid-infrared data and thus will only find objects surrounded by dusty disks (e.g., \citealt{luhman08}) or because they are carried out using narrow-band methane imaging, which probes only a specific range of spectral types (e.g., \citealt{haisch10}).

SONYC -- Substellar Objects in Nearby Young Clusters -- is
an ongoing project to provide a complete census of the brown
dwarf and planetary-mass object (PMO) population in nearby
young clusters and to establish the frequency of substellar mass
objects as a function of cluster environment. 
The survey is based on our own extremely deep optical- and near-infrared imaging,
combined with the \textit{Two Micron All Sky Survey} (2MASS) and
$Spitzer$ photometry catalogs, which are combined to create catalogs of
substellar candidates and used to identify targets for extensive spectroscopic follow-up.
Our observations are designed to reach mass limits of $0.005\,M_{\odot}$ and below, with the specific goal of probing the bottom of the IMF. 
So far, we have published results for three regions: NGC$\,$1333 \citep{scholz09a, scholz11}, $\rho$ Ophiuchi \citep{geers11}, and Chamaeleon-I \citep{muzic11}. 

$\rho$ Oph is one of the closest regions with active star formation. Its distance is estimated to be $(125 \pm 25)$ pc \citep{degeus89}. The main cloud, L1688, is a
dense molecular core, with visual extinction up to 50 -- 100 mag
\citep{wilking&lada83}. It hosts an embedded infrared cluster of
around 200 stars, inferred to have a median age of 0.3 Myr, and
surrounded by multiple clusters of young stars with a median
age of 2.1 Myr (\citealt{wilking05} and references therein).
Currently, there are $\sim 25$ spectroscopically confirmed BDs in $\rho$ Oph \citep{wilking08, alvesdeoliveira10, marsh10, geers11}. 
The current census is, however, likely incomplete due to strong and variable extinction. 
When compared to the other two SONYC targets NGC1333 and Cha-I, $\rho$ Oph suffers much stronger extinction (a few tens of magnitudes vs. $A_V < 10\,$mag). $\rho$ Oph is not as compact as NGC1333, but is more compact than Cha-I. Finally, Cha-I is slightly older than the other two clusters (2-3 Myrs vs. $\lesssim 1$Myr).

In our earlier paper on $\rho$ Oph \citep{geers11} we spectroscopically confirmed a previously unknown brown dwarf, and identified a number of new candidates based on Spitzer data.
In this paper we present the results of a second round of spectroscopy, based on our photometric
catalogs and candidate selection presented in the Geers et al. paper. We have obtained about 100 new spectra
in L1688, using the multi-object spectrograph FMOS at the Subaru Telescope, and 
the integral-field spectrograph SINFONI at the ESO's Very Large Telescope (VLT).  
This paper is organized as follows. Photometry and criteria for candidate selection are explained 
in Section~\ref{s2}, and the spectroscopic follow-up in Section~\ref{spectroscopy}. The results are 
presented in Section~\ref{specanalysis} and
discussed in Section~\ref{discuss}. Conclusions are summarized in Section~\ref{concl}.

\section{Multi-band photometry and selection of candidates}
\label{s2}

\begin{figure*}
\center
\includegraphics[width=18cm,angle=0]{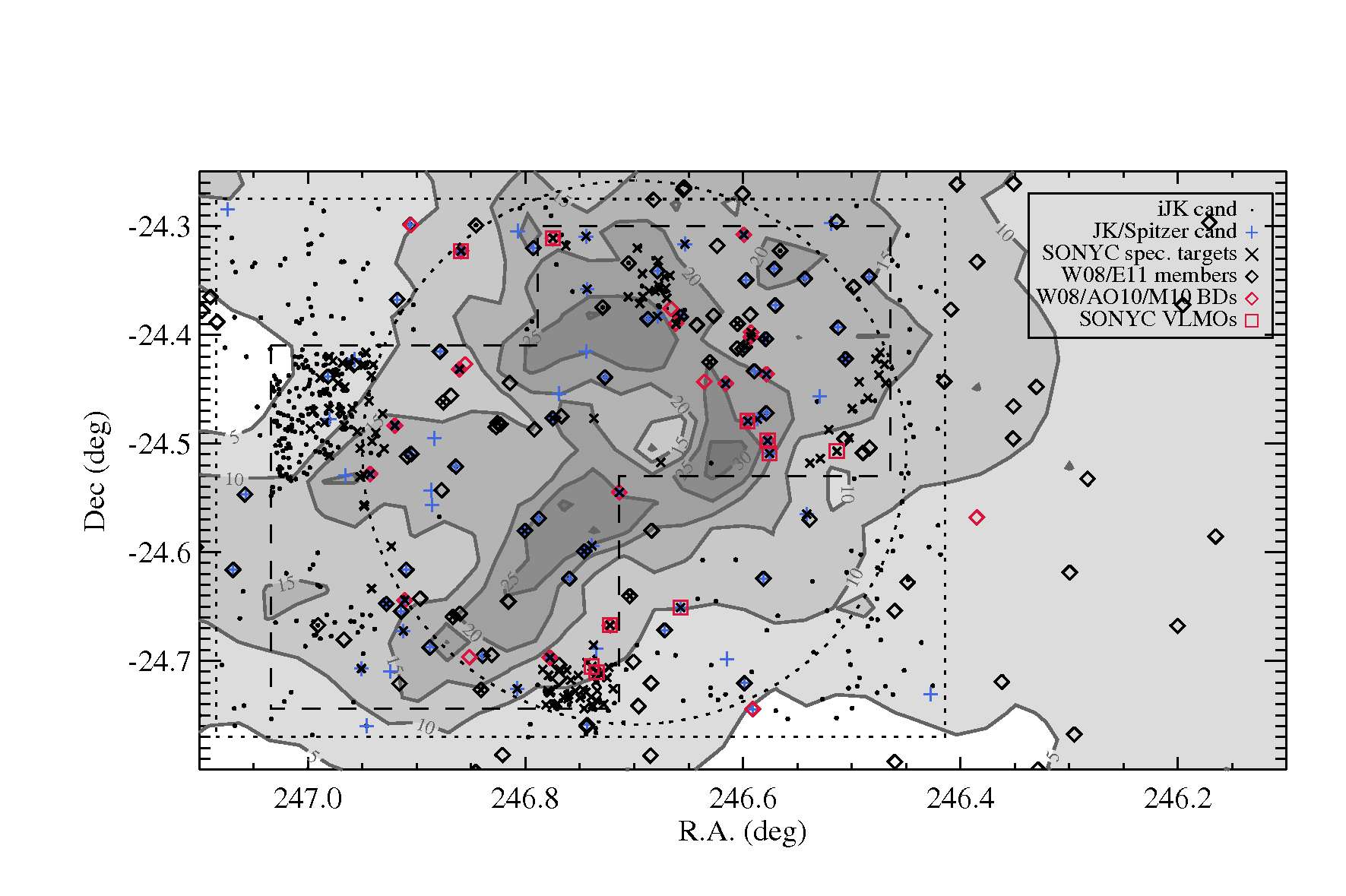}
\caption{Spatial distribution of sources in $\rho$ Oph. Contours are A$_V$ = 5, 10, 15, 20, 25, 30, as derived from 2MASS by the COMPLETE project (\citealt{ridge06}; available at www.cfa.harvard.edu/COMPLETE). 
Dotted line: $i'$-band imaging coverage; 
dashed line: $J$- and $K_S$-band imaging coverage; the circle shows the FMOS field-of-view. 
Candidate members are indicated with dots ($i'JK$ selection), and pluses ($JK/Spitzer$ selection). SONYC spectroscopic
targets (\citealt{geers11} and this work) are marked with crosses. 
$\rho$ Oph members from \citet{wilking08} and \citet{erickson11} are shown as black diamonds, while the known
BDs \citep{wilking08, alvesdeoliveira10, marsh10} are shown as red diamonds.
Red squares mark the BDs and VLM stars found in the framework of SONYC.}
\label{spatial}
\end{figure*}

\begin{figure}
\centering
\includegraphics[width=9cm,angle=0]{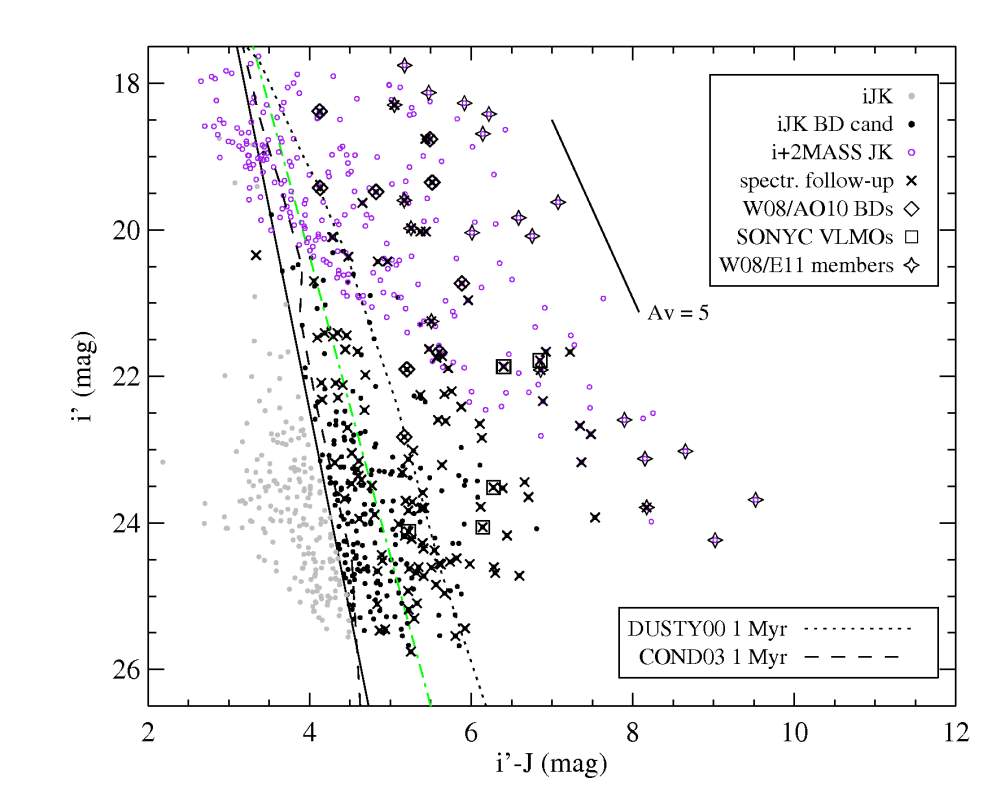}
\caption{($i'$,$i' - J$) color-magnitude diagram. Filled circles represent all sources
from our $i'JK$ catalog (MOIRCS $JK$), with dark symbols indicating the BD candidates. Sources from the 2MASS-based
$i'JK$ catalog are indicated with purple open circles. Crosses show SONYC spectroscopy follow-up targets, with spectroscopically confirmed VLMOs marked with squares. Among SONYC VLMOs, the object found at $(i' - J, i')=(5.2, 24.1)$ is identified as a low-mass star, while the remaining five are BDs. Spectroscopically confirmed brown dwarfs from the literature \citep{wilking08, alvesdeoliveira10} are shown as
diamonds, while stars indicate higher mass members of $\rho$~Oph from \citet{wilking08} and \citet{erickson11}. Atmosphere model
isochrones for age 1 Myr are overplotted for DUSTY00 \citep{chabrier00} and COND03 \citep{baraffe03}. The solid
lines denotes the BD candidate selection cutoff. The green dash-dotted line denotes the new selection cutoff (see Section~\ref{missing} for explanation). 
 }
\label{IJcmd}
\end{figure}

Observations, data reduction, as well as the cross-correlation of the optical, near-infrared and Spitzer catalogs, 
are described in detail in \citet{geers11}.
Here we give a summary of the facilities and filters used for the observations, and the description of the 
photometric catalogs used to select candidates for the follow-up spectroscopy.

The optical images in the SDSS $i'$ filter were obtained with the Subaru Prime Focus
Camera (Suprime-Cam) wide field imager \citep{miyazaki02}. The Multi-Object Infrared Camera and Spectrograph
(MOIRCS; \citealt{suzuki08}) at the Subaru telescope was used to observe $\rho$ Oph in the $J$- and $K_S$-bands.
To increase the dynamic range of the survey in the near-infrared,
we retrieved additional $JK_S$ photometry from the
2MASS point-source catalog \citep{skrutskie06}. We also retrieved the Spitzer IRAC and MIPS photometry of $\rho$
Oph from the Legacy Program data archive available at the
Spitzer Science Center, using the “High reliability catalog”
(HREL) created by the “Cores to Disks” ($c2d$) Legacy team. 
The $3.5 \sigma$ detection limit of our survey is at $i'= 26$, $J=22$, and $K_S=19.5$ mag. The completeness limits 
correspond to $i'= 24.15 \pm 0.3$, $J=20.6 \pm 0.3$, and $K_S=17.8 \pm 0.3$ mag.

To identify candidates for spectroscopy, we use two different selection procedures:\\
(1) We start with the catalog of common sources appearing in $i$, $J$, and $K_S$ catalogs. 
Low-mass and BD candidates are selected as objects detected in all three bands, and 
exhibiting a color excess in the ($i'$, $i'-J$) color-magnitude diagram. These are objects
located to the right of the candidate selection cutoff line in Figure~\ref{IJcmd}. This $iJK_S$ selection results in a catalog of 537 candidate sources
shown as dots in Figure~\ref{spatial}. 309 of these sources come from the MOIRCS $JK_S$ catalog (black circles in Figure~\ref{IJcmd}), and the rest
from 2MASS (purple circles); and \\
(2) We cross-match $JK_S$ catalogs with the $Spitzer$ catalog and select objects with IRAC colors consistent with or redder than sources with a circumstellar disk (Figure~\ref{spitzerplot}). $JK/Spitzer$ selection yields 83
candidate members of $\rho$ Oph, within the boundaries of our optical survey. These objects
are marked with '+' in Figure~\ref{spatial}. 26 of these are found to be BD candidates based on their near-infrared 
magnitude and color, with 12/26 designated as spectroscopically confirmed BDs in the literature.

35 out of 83 candidates from the $JK/Spitzer$ list have an optical counterpart. 33/35 are 
also identified as candidates in the $iJK_S$ selection, while the remaining two are located to the left of the candidate selection line 
in Figure~\ref{IJcmd}. We obtained a spectrum for one of these two sources, and found that it is probably not a member of $\rho$ Oph.
We note that the majority (32) of the 35 overlapping $JK/Spitzer$ candidates are part of the 2MASS-based $iJK_S$ catalog, and only 3 are from the 
MOIRCS $iJK_S$ candidate selection.


\section{Follow-up spectroscopy}
\label{spectroscopy}
In our first spectroscopy run in May 2009, we used MOIRCS/Subaru to obtain the spectra of 58 BD candidates
selected from our $iJK_S$ catalog, and one previously known BD candidate member \citep{geers11}. The 58 selected
candidates all come from the $iJK_S$ catalog based on the MOIRCS $JK_S$ photometry (309 objects).
During 2010 we were awarded time at the Subaru Telescope and the VLT that allowed us to obtain spectra of $\sim$100  additional substellar candidates in $\rho$ Oph. These new observations, data reduction and number statistics of the observed candidates from the two catalogs are presented in this section.
Candidates with spectroscopic follow-up are marked with crosses in Figure~\ref{IJcmd} and squares in Figure~\ref{spitzerplot}.

\subsection{FMOS/Subaru}
We obtained spectra in the near-infrared $J$- and $H$-bands (0.9 -- 1.8 $\mu$m) using the Fiber Multi Objects Spectrograph (FMOS) at the
Subaru Telescope \citep{kimura10}. The fiber positioner system of FMOS configures 400 fibers in the 30-arcmin diameter field of view, and the spectra are extracted by the two spectrographs (IRS1 and IRS2). Our data were obtained during the S10A semester in a shared-risk mode, using only one of the spectrographs (IRS1; 200 fibers) in the low-resolution mode 
(R~$\sim600$).
Data reduction was carried out using the software package supplied by Subaru during the observing run. The package consists of IRAF tasks and C programs using the CFITSIO library. 
The recommended telluric standards for FMOS are stars of spectral type F, G or early K, at a known (and preferably low) extinction. As the stars in our science field suffer strong extinction, we observed an additional 
field at the same average airmass as $\rho$ Oph, containing several G0V stars that were used for telluric correction. 

We observed $\rho$ Oph during the two nights of June $3 - 4$ 2010, and obtained in total 81 spectra. 14 of the objects came from the $JK/Spitzer$ candidate list, 46 from the $iJK_S$-based candidate selection, and the remaining fibers were placed on a number of known members of spectral type K and M, to serve as a reference. We note that the $iJK_S$ candidate
selection is based only on the MOIRCS $JK_S$ catalog (309 objects).

\subsection{SINFONI/VLT}
The spectra of 13 objects from the $JK/Spitzer$ candidate list were obtained using the integral-field spectrograph SINFONI at the ESO/VLT, in the $H$- and $K$-bands (program ID 385.C-0450). Two of the objects have also been observed with FMOS. 
The observations were carried out one object at a time, using the largest available FOV offered by SINFONI ($8''\times8''$) and in the no-AO mode. The spectral resolution delivered by the H+K setting of the instrument is R~$\sim 1500$. Data reduction was performed using the SINFONI pipeline supplied at ESO. Telluric correction was performed with the help of standard stars observed before and after our targets. Standard stars were all of the early B-type, which exhibit a large number of hydrogen lines throughout the $H$- and $K$-bands. Some of the lines can be manually removed from the standard-star spectra (e.g. Br$\,\gamma$, Br$\,$10, Br$\,$11), while others appear blended with the telluric features and thus cannot be easily removed. Some residuals due to uncorrected hydrogen lines appear in our spectra, but do not affect the results, as we are interested only in broad features. These residuals mainly mimic emission lines in the 1.55 -- 1.65 $\mu$m region in some of the SINFONI spectra.

\section{Spectral analysis}
\label{specanalysis}

\subsection{Spectral fitting procedure}
\label{fitting}
The spectral analysis is identical to the one presented in \citet{scholz09a} and \citet{geers11}.
Spectra of young objects later than M5 show a characteristic broad peak in the $H$-band \citep{cushing05}.
The feature is caused by water absorption on both sides of the $H$-band and its depth depends strongly on effective temperature. While the $H$-band peak appears round in old field dwarfs, it is unambiguously
triangular in young, low-gravity sources \citep[e.g.][]{brandeker06, kirkpatrick06}.
In addition, young BDs are expected to have
flat or increasing $K_S$-band spectra with CO absorption bands at
$\lambda > 2.3 \mu$m, and a sharp drop in flux at the edge of the $J$-band ($\sim 1.3\mu$m), again due to 
water absorption.

The spectra were first assessed through visual inspection.
Very noisy spectra ($\sim 5 \%$ of all spectra) and those lacking the above mentioned features characteristic for BDs 
($\sim 75 \%$ of all spectra) are discarded. 
The latter sources exhibit smooth, featureless spectra, which make it difficult to constrain their nature.  
The majority of these objects are expected to be background late-type giant or dwarf stars. Some of them could be embedded YSOs with spectral types earlier than M, although we do not
have any evidence for youth. This selection yields 21 VLM candidates.

In a second step, we estimate the effective temperatures of the selected
VLM candidates by fitting model spectra from the DUSTY series \citep{allard01}, 
following the procedure outlined in \citet{scholz09a}. 
Since
extinction and effective temperature are difficult to determine
separately with low-resolution, low signal-to-noise
spectra, we started by adopting the $A_V$ as determined from
the $J-K_S$ colors, assuming an intrinsic BD photospheric $J-K_S$ color of 1 mag.
We apply the extinction law from \citet{cardelli89}, assuming $R_V\,$=$\,$4. 
We vary the effective temperature from 1800~K to 3900~K, in steps of 100~K, keeping
the extinction at a fixed value. The best fit temperature is determined from a $\chi^2$ analysis. 
Although in almost all cases this procedure results in a decent match between the 
observed spectra and the best-fit models, we note that small adjustments to the value of the extinction
($\Delta A_V \le 2$) can slightly improve the quality of the fit. 
This adjustment is determined by changing $A_V$ in steps of 1 around the 
value derived from the $J-K_S$ color, rounded to the closest integer. 
For 21 out of 23 entries in Table~\ref{tspec}
either a zero or a small adjustment (up to $A_V$=2) is required, while in the remaining two cases $\Delta A_V \approx 3$ 
results in a better fit. 
Both values of $A_V$ are listed in Table~\ref{tspec}.  

As demonstrated in \citet{scholz09a}, it is often difficult to reliably distinguish between two 
models with $\Delta T = 100 K$. However, for $\Delta T \gtrsim 200 K$ there are discernible differences between
models and observations, combined with a clear increase in $\chi^2$. Thus, a conservative estimate
for a typical uncertainty in $T_{\mathrm{eff}}$ is $\pm 200\,$K. In cases where an equally good fit is obtained
using two models separated by 100~K, as the best-fit $T_{\mathrm{eff}}$ in Table~\ref{tspec} we list the average value, while  
in Figures~\ref{specFMOS} and \ref{specSINFO} we plot the model with the lower $T_{\mathrm{eff}}$. 
Although the temperature fitting grid extends up to 3900~K, 
we note that our fitting scheme does not work that well above 3200~K. The reason is the shape of
the H-band
peak that becomes flat, making it difficult to discriminate between the models at higher temperatures. 
The situation is further complicated by high extinctions typically encountered in the $\rho$ Oph region. 
We find spectra of three objects to be consistent with effective temperatures of 3200~K or somewhat above. We set 3200~K
as a lower temperature limit for these objects in Table~\ref{tspec}.

\subsection{Results of the spectral fitting}
\label{results}

\begin{figure*}
\center
\includegraphics[width=16cm,angle=0]{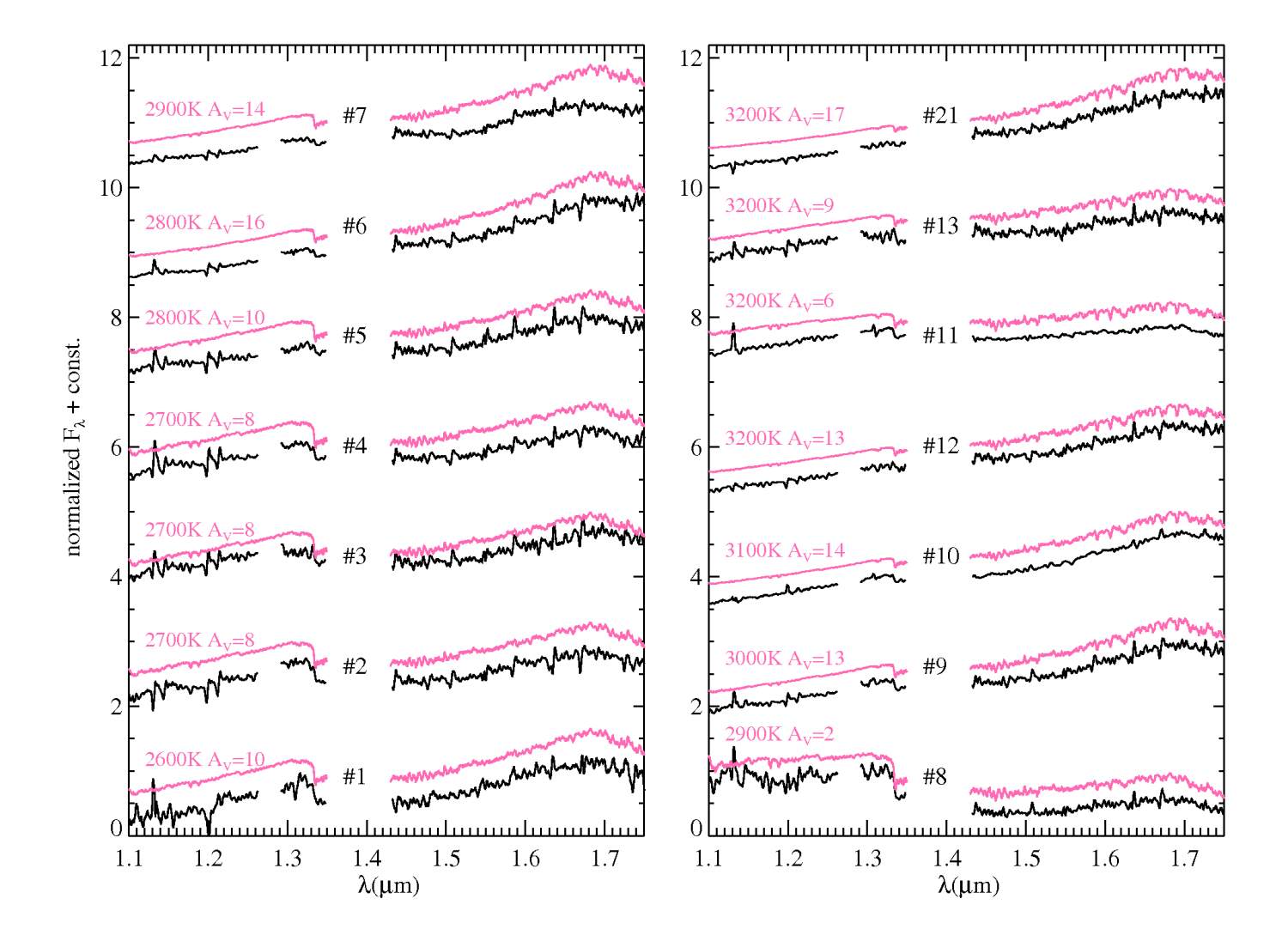}
\caption{Spectra of very-low-mass objects in $\rho$ Oph, observed with FMOS/Subaru. The spectra
are shown in black, with the best-fit models (AMES-Dusty; \citealt{allard01}) in color. 
The models are slightly offset for clarity. 
The spectra and the models were smoothed prior to plotting, so that the quality of the fit 
to the broad-band features can be appreciated. 
The models have been reddened by the best-fit A$_V$, given in Table~\ref{tspec}.
The spectra are plotted in the order of increasing $T_{\mathrm{eff}}$ from bottom to top and from left to right.}
\label{specFMOS}
\end{figure*}

\begin{figure*}
\center
\includegraphics[width=16cm,angle=0]{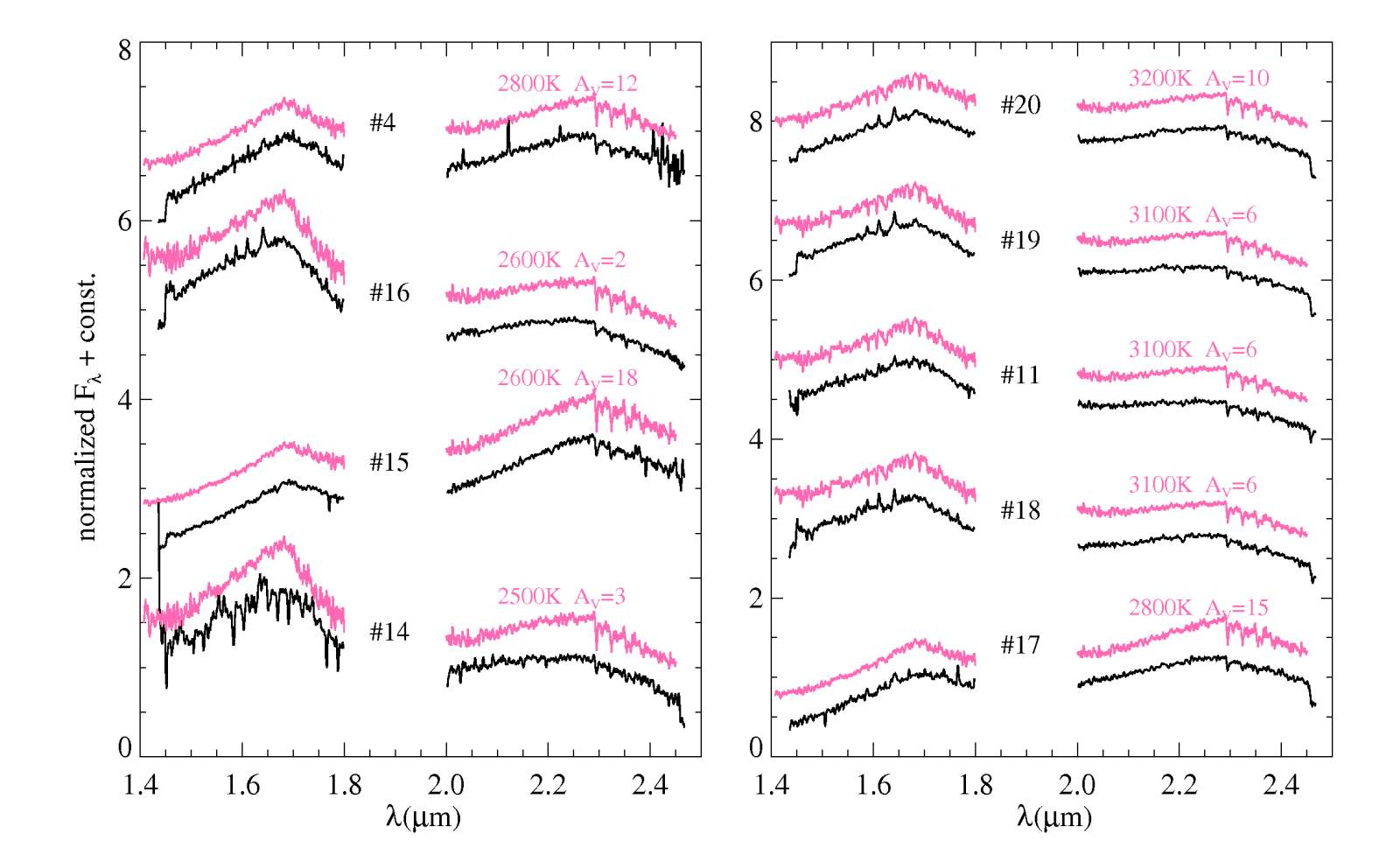}
\caption{Spectra of very-low-mass objects in $\rho$ Oph, observed with SINFONI/VLT. See caption
to Figure~\ref{specFMOS} for plot details.}
\label{specSINFO}
\end{figure*}

\begin{figure}
\center
\includegraphics[width=8cm,angle=0]{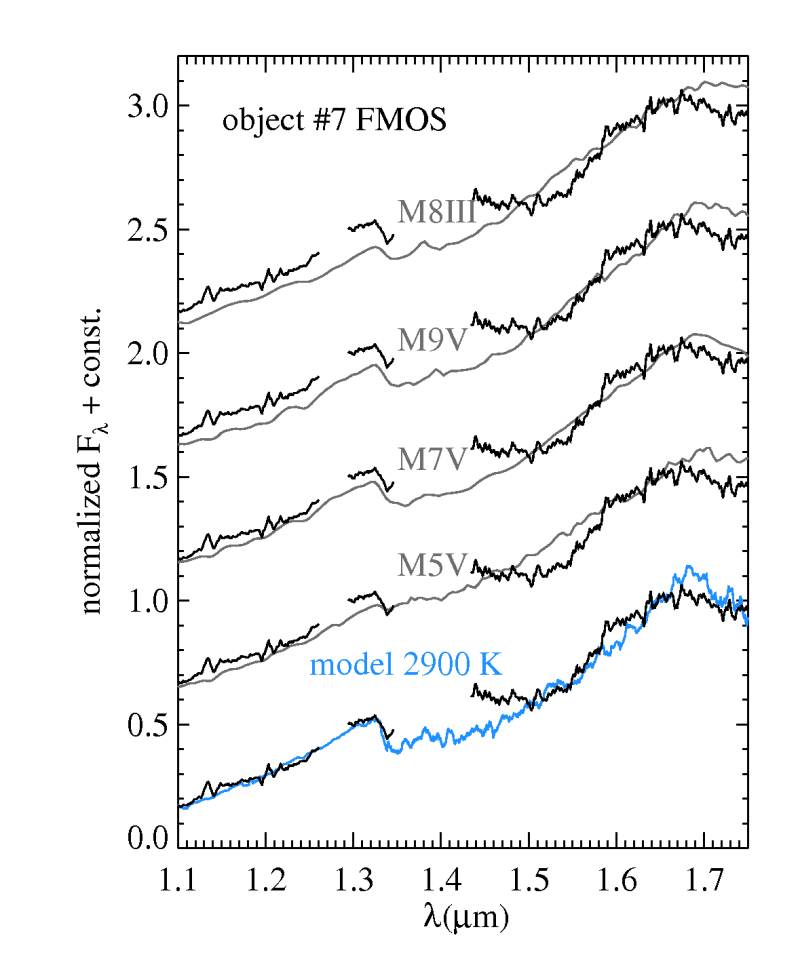}
\caption{Spectrum of the object \#7 (black; SONYC-RhoOph-5) in comparison with the 
best-fit low-gravity model (blue) and spectra of three field dwarfs of spectral type M5 (2MASSJ01532750+3631482; \citealt{burgasser04}), 
M7 (VB 8; \citealt{burgasser08}) and M9 (2MASS J12425052+2357231 \citealt{kirkpatrick10}), as well as 
a spectrum of an M8 giant (HD 113285; SpeX Libraries) shown in grey. The model and template spectra have been reddened by $A_V=14$ and normalized at $1.65\mu$m.}
\label{no7}
\end{figure}

The spectra of 21 BDs and very-low-mass objects (VLMO) are shown in Figure~\ref{specFMOS} (FMOS) and Figure~\ref{specSINFO} (SINFONI). 
Two objects appear in both datasets ($\#$4 and $\#$11), 
The best fit $T_{\mathrm{eff}}$ resulting from the two independent fits
are consistent, differing by only 50 K and 100 K, respectively. 
Spectra are shown in black and the best-fit models in color. For clarity, the models are slightly 
offset from the spectra. Both models and spectra have been smoothed with the boxcar function of 30~\AA~
 prior to plotting, to make it
easier to appreciate the quality of the broad-band fit. 
The 21 BDs and VLMOs with SINFONI and/or FMOS spectra are listed in Table~\ref{tspec}, including the 
best model fit parameters. 

Our observations result in 13 probable substellar members (defined here as objects with $T_{\mathrm{eff}} \leq 3000$K) 
and 8 very-low-mass stars (3000 K $< T_{\mathrm{eff}} \leq 3200$K). 
Among these objects, we report 6 new likely BDs and 2 new very-low-mass stars.
We label those objects SONYC-RhoOph-3 to 10, following the nomenclature of \citet{geers11} who published the first two discoveries. 
The six new BDs have effective temperatures between 2600~K and 3000~K, which, according to the 1Myr
DUSTY and COND isochrones, correspond to the masses between 0.03$\,$M$_{\odot}$ and 0.1$\,$M$_{\odot}$.

The H-band spectrum of the object $\#7$ (SONYC-RhoOph-5) appears to have a slightly
rounded shape when compared
to the spectra of other objects in Figure~$\ref{specFMOS}$. However, neither the comparison with the spectra of field
dwarfs or giants provides a satisfactory match with the observed spectrum (see Figure~\ref{no7}).
The source is detected in all four IRAC bands and shows a clear mid-IR excess: with I$1-$I2 $=0.41$ and I$3-$I4 $=1.03$, 
it resides within the class-II object area in Figure~\ref{spitzerplot}.  
We thus conclude that SONYC-RhoOph-5 is a probable new BD member of the cluster.

The H-band peak around 1.68$\mu$m in the spectrum of the object $\#14$ might not appear as sharp as
expected for a young BD with $T_{\mathrm{eff}}=2550$K, but this can be attributed to the strong noise
in this spectrum. In \citet{alvesdeoliveira10} one can clearly see that this source exhibits the sharp H-band peak
characteristic for young BDs. They classify the source as M8.25, which is consistent both with the
effective temperature obtained from our fits, and the spectral type derived from various indices (see Section~\ref{S_SpT} and Table~\ref{T_SpT}). 

In this paper and in previous papers of the SONYC series, we have chosen to draw a line between the
stellar and substellar regime either at $T_{\mathrm{eff}}=3000$~K or at SpT$\,$=$\,$M6. 
According to the COND03 and DUSTY models, a 1 Myr old object at the substellar boundary will have 
an effective temperature of about 3000~K. 
In the scale defined by \citet{luhman03}, this limit corresponds to a spectral type M6. However, the relation
between the effective temperature and spectral type for young VLMOs is not that well established. 
There are two objects that have at least once been classified as brown dwarfs in the literature (M6 or later) and that
we do not define as substellar. The first one is the source $\#21$ which we find to have $T_{\mathrm{eff}} \geq 3200$K. 
\citet{wilking99} and 
\citet{luhman&rieke99} classify this source as M7 and M6.5, respectively.
Based on our fits, we identify this source a very-low-mass star, rather than a BD. 

The second object  where we find discrepancy with the literature classification is found at 
($\alpha$, $\delta$)$_{2000}$=(16:26:31.36, -24:25:30.2) and was classified as M6 by \citet{wilking99} from the NIR data, but
also as M5 (NIR; \citealt{luhman&rieke99}) and M3.75 (optical, \citealt{wilking05}). We find the H-band spectrum to be
much flatter than in BD spectra, resulting in $T_{\mathrm{eff}}$ clearly above 3200~K. Thus, we discard 
this object as a potential BD, and conclude that our spectrum is  
more consistent with the earlier spectral classification from the literature.
 
\begin{deluxetable*}{lllcccclll}
\tabletypesize{\scriptsize}
\tablecaption{Parameters of $\rho$ Oph M-type objects observed with FMOS/Subaru and SINFONI/VLT \label{tspec}}
\tablewidth{0pt}
\tablehead{\colhead{\#} & 
	   \colhead{$\alpha$(J2000)} & 
	   \colhead{$\delta$(J2000)} & 
	   \colhead{ins\tablenotemark{a}} & 
	   \colhead{A$_{V}$\tablenotemark{b}} & 
	   \colhead{A$_{V}$\tablenotemark{c}} & 
	   \colhead{$T_{\mathrm{eff}}$ (K)} & 
	   \colhead{SpT} & 
	   \colhead{Ref\tablenotemark{d}}	&
	   \colhead{name(s)\tablenotemark{e}}}
\tablecolumns{10}
\startdata
1  & 16 26 03.28 & -24 30 25.8 & F & 8.4  & 10 & 2650 &	\nodata		& \nodata	& \scriptsize{SONYC-RhoOph-7} \\ 
2  & 16 27 05.93 & -24 18 40.2 & F & 7.8  & 8  & 2700 & \nodata		& \nodata	& \scriptsize{SONYC-RhoOph-6} \\ 
3  & 16 27 38.63 & -24 38 39.2 & F & 6.6  & 8  & 2750 & M8.5, M7, M6	& 1$\,$2$\,$3 	& \scriptsize{GY310}\\	  	 
4  & 16 26 22.27 & -24 24 07.1 & F & 9.7  & 8  & 2750 & M6.5, M8.5	& 1$\,$3	& \scriptsize{GY11}\\ 		 
4  & 16 26 22.27 & -24 24 07.1 & S & 9.7  & 12 & 2800 & M6.5, M8.5	& 1$\,$3	& \scriptsize{GY11}\\	    	 
5  & 16 26 18.82 & -24 26 10.5 & F & 9.4  & 10 & 2850 & M7.5, M5.5, M7	& 1$\,$2$\,$3	& \scriptsize{CRBR 14; ISO-Oph-23}\\
6  & 16 26 18.98 & -24 24 14.3 & F & 17.6 & 16 & 2850 & M5		& 1$\,$2	& \scriptsize{CRBR 15}\\    	 
7  & 16 26 22.96 & -24 28 46.1 & F & 13.2 & 14 & 2900 & \nodata 	& \nodata	& \scriptsize{SONYC-RhoOph-5}\\  
8  & 16 26 51.28 & -24 32 42.0 & F & 2.3  & 2  & 2900 & M8, M8.5	& 5$\,$4	& \scriptsize{GY141}\\ 	 	 
9  & 16 26 53.35 & -24 40 02.3 & F & 12.4 & 13 & 3000 & \nodata		& \nodata	& \scriptsize{SONYC-RhoOph-3}\\  
10 & 16 26 22.19 & -24 23 52.4 & F & 13.8 & 14 & 3100 & M8.5, M6.5	& 1$\,$2	& \scriptsize{GY10}\\  	  	 
11 & 16 26 27.81 & -24 26 41.8 & F & 6.4  & 6  & $\geq$3200 & M5, M6		& 6$\,$1	& \scriptsize{GY37}\\		 
11 & 16 26 27.81 & -24 26 41.8 & S & 6.4  & 6  & 3100 & M5, M6 		& 6$\,$1	& \scriptsize{GY37}\\ 		 
12 & 16 26 18.14 & -24 30 33.0 & F & 14.7 & 13 & 3200 & \nodata		& \nodata	& \scriptsize{SONYC-RhoOph-4}\\  
13 & 16 27 46.29 & -24 31 41.2 & F & 8.3  & 9  & $\geq$3200 & M6	& 3		& \scriptsize{GY350}\\		 
14 & 16 27 40.84 & -24 29 00.7 & S & 3.1  & 3  & 2550 & M8.25		& 7		& \scriptsize{CFHTWIR-Oph96}\\	 
15 & 16 26 18.58 & -24 29 51.4 & S & 15.0 & 18 & 2600 & \nodata		& \nodata	& \scriptsize{SONYC-RhoOph-8, CFHTWIR-Oph16}\\      
16 & 16 27 26.58 & -24 25 54.4 & S & 0.9  & 2  & 2650 & M8		& 6		& \scriptsize{GY264}\\			    	    
17 & 16 27 26.22 & -24 19 23.0 & S & 13.6 & 15 & 2800 & \nodata		& \nodata	& \scriptsize{SONYC-RhoOph-10; CFHTWIR-Oph78}\\	    
18 & 16 26 37.81 & -24 39 03.2 & S & 6.2  & 6  & 3100 & \nodata		& \nodata	& \scriptsize{SONYC-RhoOph-9; CFHTWIR-Oph31}\\	    
19 & 16 27 06.60 & -24 41 48.8 & S & 3.7  & 6  & 3150 & M5.5, M6	& 6$\,$3	& \scriptsize{GY204}	\\	  		    
20 & 16 26 23.81 & -24 18 29.0 & S & 8.2  & 10 & 3200 & M6.7, M7	& 5$\,$8	& \scriptsize{CRBR 31}  \\ 			    
21 & 16 27 05.98 & -24 28 36.3 & F & 14.7 & 17 & $\geq$3200 & M7, M6.5	& 1$\,$2  	& \scriptsize{GY202}
\enddata
\tablecomments{Based on AMES-Dusty models \citep{allard01}} 
\tablenotetext{a}{Instrument used for spectroscopy. F: FMOS/Subaru, S: SINFONI/VLT}
\tablenotetext{b}{Visual extinction in magnitudes, and calculated from $J-K$ color, assuming the intrinsic $J-K$ of 1.}
\tablenotetext{c}{Best-fit extinction from spectral fitting, in magnitudes.}
\tablenotetext{d}{Spectral-type references: 
[1] \citet{wilking99}; 
[2] \citet{luhman&rieke99};
[3] \citet{natta02}; 
[4] \citet{luhman97};
[5] \citet{cushing00}; 
[6] \citet{wilking05}; 
[7] \citet{alvesdeoliveira10}
[8] \citet{wilking08}.}
\tablenotetext{e}{References for identifiers:
GY: \citet{greene&young92};
CFHTWIR: \citet{alvesdeoliveira10};
CRBR: \citet{comeron93};
ISO-Oph: \citet{bontemps01}.}
\label{tspec}
\end{deluxetable*}

\subsection{Spectral type indices}
\label{S_SpT}

\begin{deluxetable}{rlll}
\tabletypesize{\scriptsize}
\tablecaption{Spectral types calculated from the SINFONI spectra}
\tablewidth{220pt}
\tablehead{
\colhead{} & \multicolumn{3}{c}{Spectral Type} \\
\colhead{\#} & \colhead{HPI\tablenotemark{a}} & \colhead{Q\tablenotemark{b}} & \colhead{H$_2$O\tablenotemark{c}}}
\tablecolumns{4}
\startdata
1 & M9.7 & \nodata & M9.6 \\
2 & M8.6 & \nodata & M6.9 \\
3 & M8.3 & \nodata & M6.6  \\
4 & M8.5, M8.0 & M6.1 & M6.6, M7.8 \\
5 & M8.5 & \nodata & M8.0 \\
6 & M8.0 & \nodata & M5.7 \\
7 & M7.4 & \nodata & M5.7 \\
8 & M8.0 & \nodata & M6.3 \\
9 & M7.5  & \nodata & M5.1 \\
10 & M7.4 & \nodata & M5.5 \\
11 & $\leq$M7, M7.1 & M5.8 & $\leq$M5, M5.8 \\
12 & M7.0 & \nodata & M5.0\\
13 & M7.1 & \nodata & $\leq$M5 \\
14 & M9.3 & M7.1 & \nodata \\
15 & M9.1 & M8.7 & M8.0 \\
16 & M8.6 & M8.0 & M8.0 \\ 
17 & M7.8 & M6.2 & M8.4 \\
18 & M7.0 & M6.8 & M5.4 \\
19 & M7.1 & M5.9 & M6.4 \\
20 & M7.0 & M6.6 & M5.7 \\
21 & $\leq$M7 & \nodata &  $\leq$M5		
\enddata
\tablenotetext{a}{Derived from the HPI index \citep{scholz11}}
\tablenotetext{b}{Derived from the Q-index \citep{wilking99,cushing00}.}
\tablenotetext{c}{Derived from the H$_2$O-index \citep{allers07}}
\label{T_SpT}
\end{deluxetable}

In this section we attempt to assign spectral types to the VLMOs from our survey, based on various
indices suggested in the literature. The SINFONI spectra are particularly suited for a comparison of different methods, 
thanks to the spectral coverage, resolution, and overall quality
of the spectra. We determine spectral types based on the following three indices: \\
(1) H-peak index (HPI) defined by \citet{scholz11} and based on the ratio between the average 
fluxes in $1.675-1.685\mu$m and $1.495-1.505\mu$m wavelength regions. The HPI takes advantage of the slope on the blue side
of the H-band peak observed in late-M and L-type objects, and is related to the spectral type as
SpT$\,= -0.84+7.66\,\cdot\,$HPI.
The index is highly sensitive to the spectral type
in the $\geq$M7 regime, but cannot be used reliably for earlier spectral types where
the slope of the feature becomes too shallow.\\
(2) Q-index defined in \citet{wilking99} as $Q=(F1/F2)(F3/F2)^{\beta}$.
F1, F2, and F3 are the average values of the relative
flux densities computed in the three bands $2.07-2.13$, $2.267-2.285$, and $2.40-2.50 \mu$m, respectively.
The useful portion of our SINFONI spectra extends only to $\sim2.45\mu$m. As evident from Figure~\ref{specSINFO}, the
continuum of the spectra longward of $\sim2.3\mu$m can be approximated by a linear function. We fit a linear function 
to the wavelength range $2.30-2.45 \mu$m ($2.43\mu$m in case of the object $\#4$) and extrapolate to the $2.45-2.50\mu$m 
region. We use $\beta=1.26$ and the $SpT-Q$ relation derived for $\rho$ Oph by \citet[][Eq.~1 and 2]{cushing00}.
\\
(3) The $H_2O$-index proposed by \citet[][Eq.~1]{allers07}. This index is defined as the ratio between the average
fluxes in $1.550-1.560\mu$m and $1.492-1.502\mu$m wavelength regions, and is linearly dependent on spectral
type between M5 and L0.\\
All spectra have been extinction corrected using the best-fit value of extinction obtained from our model
fits (see Section \ref{specanalysis}), and assuming the extinction law from \citet{cardelli89}. 
The spectral types resulting from the three indices are shown in Table~\ref{T_SpT}. 
The wavelength coverage of the FMOS data does not extend to the K-band, and thus
they cannot be used
to calculate the Q-index.
For the object $\#14$, one of the narrow band regions used to calculate the $H_2O$-index coincides with 
a strong noise feature, which prevent us to use this index to derive a reliable spectral type. 
For the objects $\#4$ and $\#11$, we list spectral type estimates based on their FMOS and SINFONI spectra, in that order.  

We can make a comparison between the spectral types derived from different indices, excluding those
where we only list the limits on the spectral type. 
The quoted uncertainties for the spectral types derived from the three indices are $\pm0.4$ for HPI \citep{scholz11}, $\pm1.0$ for the H$_2$O-index \citep{allers07}, and $\pm1.5$ for the Q-index \citep{wilking99}.
When comparing the HPI-derived values with those based on the other
two indices, we find that our spectral type scheme gives systematically later spectral types (in all cases except one). 
On average, the difference between the HPI and H$_2$O spectral types is 1.3. The difference is more pronounced 
for the FMOS data: the average value is 1.6, while for the values derived from the SINFONI data we get better
agreement (0.9 spectral types). This can probably be attributed to the better spectral resolution and signal-to-noise 
of the SINFONI data.
The average difference between the HPI and Q-index spectral types is 1.1. 

We tested the effect that uncertainties in the fitted $A_V$ have on the derived spectral types.
A change in the extinction of $\Delta A_V = 2$ results in $\Delta SpT \sim 0.6 $ subtypes, for the spectral types
derived from the HPI and the $H_2O$-index, while the Q-index is esentially reddening-independent. The change
in the $A_V$ affects the HPI and $H_2O$ spectral types for our sample objects in a similar way. 
A possible explanation for the discrepancies between spectral types derived from the HPI and other two indices
might be in the extinctions of the sample from which HPI was derived. HPI
was derived for objects with low extinction, i.e. the extinction correction was less problematic. 
 The majority of our objects in $\rho$ Oph
are at large extinctions, which will introduce additional uncertainty to the spectral type as derived from the HPI.
On the other hand, the sample that was used to derive the $H_2O$ index covers a larger range of extinctions.

Given the uncertainties, we conclude that the spectral types derived from the different indices are in agreement, with
the HPI spectral types being systematically later by about one spectral subtype from the values derived from the other
two indices.

\section{Discussion}
\label{discuss}

\subsection{Objects with mid-infrared color excess}
\label{S_MIR}

\begin{figure*}
\center
\includegraphics[width=15cm,angle=0]{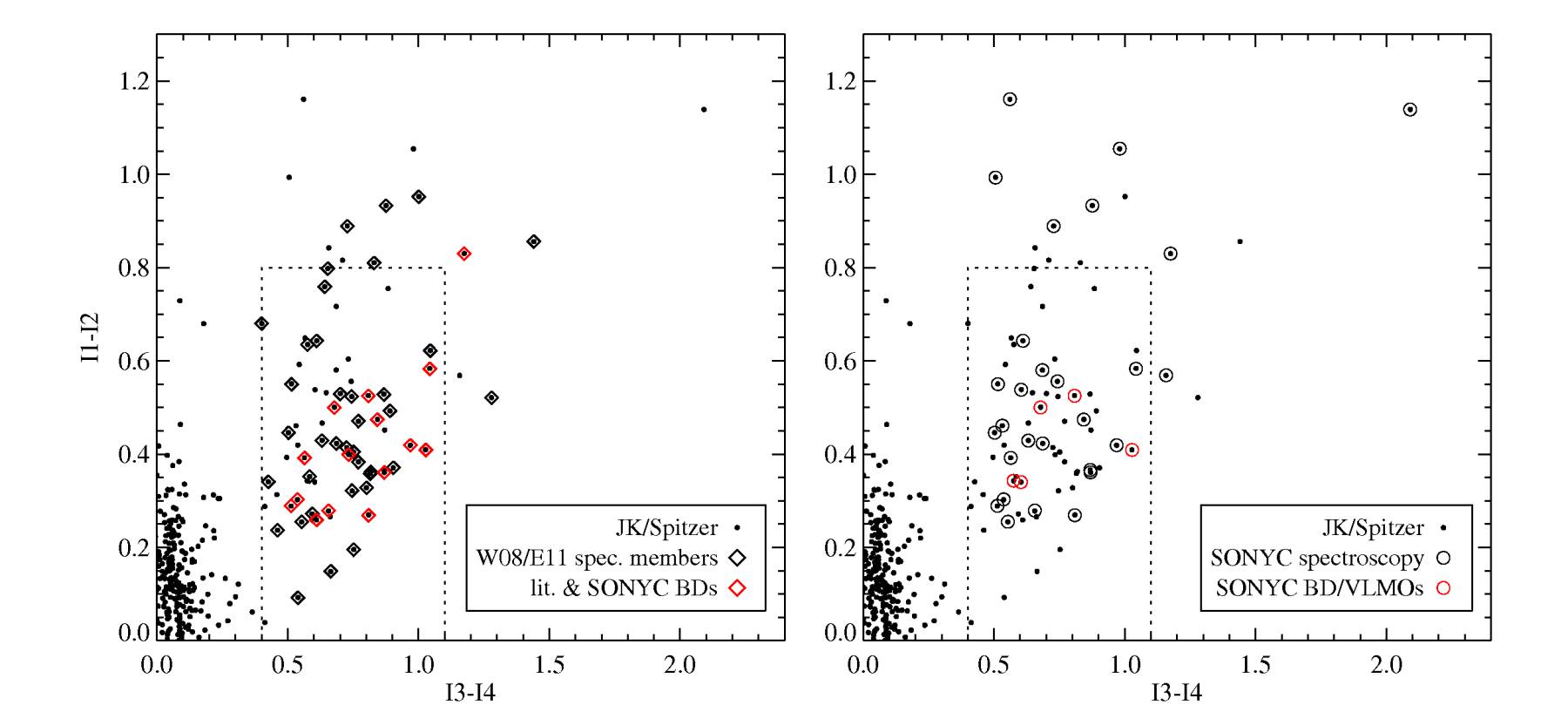}
\caption{{\it Left:} Color - color diagram in ([3.6] -- [4.5], [5.8] -- [8.0]), constructed from Spitzer
IRAC photometry, for 2MASS and MOIRCS sources with
a Spitzer match (black dots). The dashed line denotes the area where class II objects are
located, based on \citet{allen04}. Spectroscopically confirmed members of $\rho$ Oph from \citet{wilking08} and \citet{erickson11} are represented
by diamonds, with the confirmed brown dwarfs (defined as the objects with spectral type $\geq\,$M6, from \citealt{wilking08, alvesdeoliveira10} and this work) shown in red. 
{\it Right:} Same as left; open circles mark the objects with spectra obtained as part of SONYC (five using MOIRCS/MOS, and the rest using FMOS and SINFONI). SONYC BDs and VLMOs are shown as the red symbols.}
\label{spitzerplot}
\end{figure*}

\begin{figure}
\center
\includegraphics[width=9cm,angle=0]{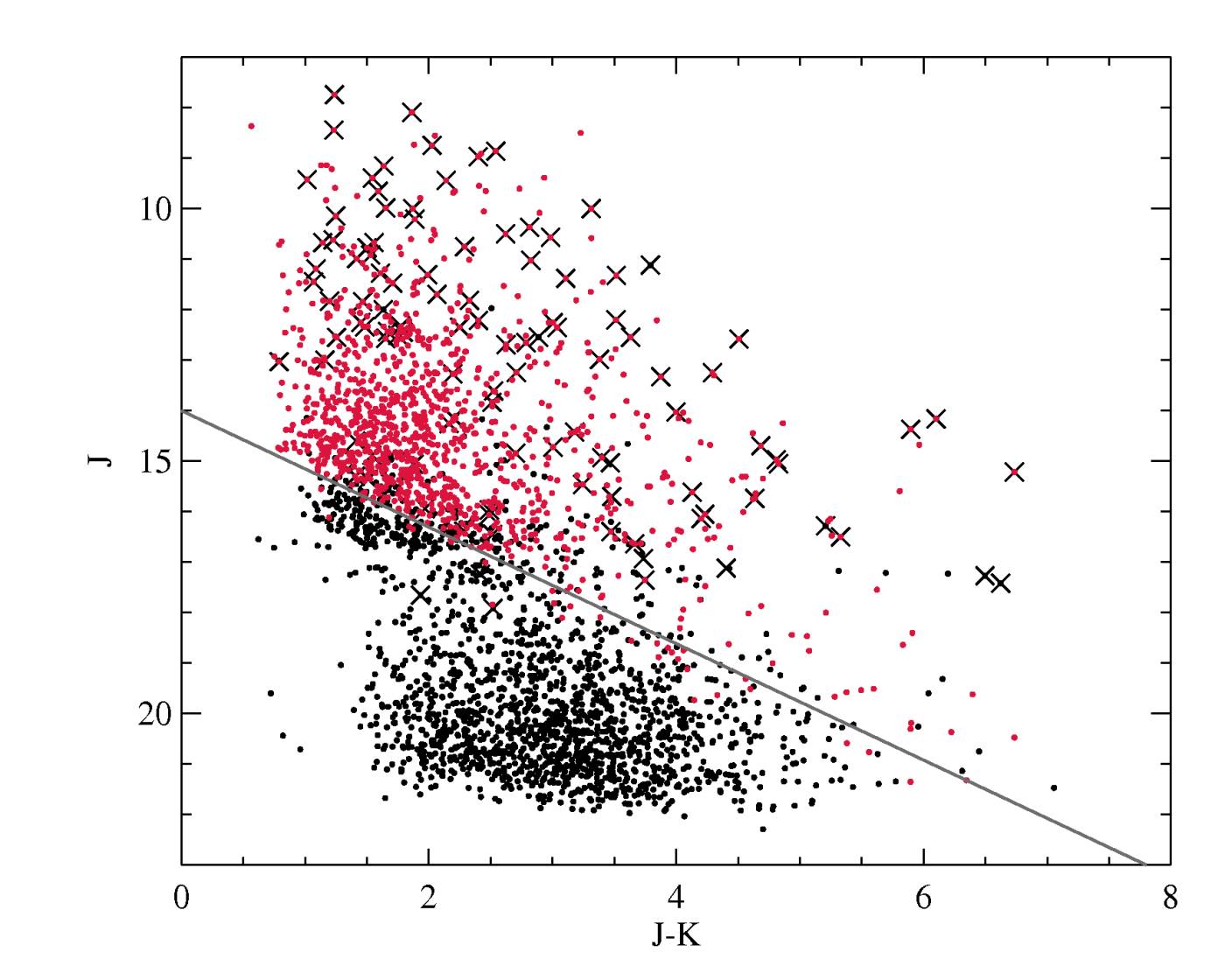}
\caption{Near-infrared color-magnitude diagram for the sources in $\rho$ Oph, based on MOIRCS and 2MASS photometry.
Spectroscopically confirmed stellar and substellar members (\citealt{wilking08}, \citealt{erickson11}, \citealt{alvesdeoliveira10}, and this work) are marked with crosses; objects with Spitzer counterpart are shown in red. The solid
line shows the approximate limit of the combined $JK-Spitzer$ catalog.}
\label{JKcmd}
\end{figure}

Figure \ref{spitzerplot} shows a color-color diagram constructed from Spitzer IRAC photometry. 
The dashed line denotes the area where class II objects are
located, based on \citet{allen04}.
For 83 objects, the IRAC colors are consistent with
or redder than sources with a circumstellar disk, here
defined as (I3 - I4) $>0.4$ and (I1 - I2) $>0.0$. 
46 of these objects appear in the list of spectroscopically
confirmed $\rho$ Oph members (diamonds; \citealt{wilking08}).
33 objects have been observed in the framework of SONYC (black squares). 
The two samples have 19 objects in common.
There are 12 spectroscopically confirmed brown dwarfs
from \citet{wilking08} and \citet{alvesdeoliveira10} and 3 identified in this work 
appearing in this plot. There are 18 objects with Spitzer excess that, to our knowledge, do not 
have available spectra (objects shown only as crosses to the right of (I3 - I4) = 0.4 in Figure~\ref{spitzerplot}).

Of 83 candidate members with mid-infrared excess, 65 have been either observed by us, or have been
spectroscopically confirmed as members in the literature. 15 of them have spectral type 
later than M5 or estimated effective temperature at or below 3000~K. 
Thus, we conclude that the number of missing brown dwarfs in this sample is $ (83-65)\times 15/65 \approx 4$. Of course, the total number of 
missing substellar objects in $\rho$ Oph is probably much higher. Our estimate is valid only for the 
objects with disks (i.e. exhibiting MIR excess), and down to a certain limit in mass and extinction.
This limit is primarily imposed by the sensitivity of the Spitzer data. The solid line in Figure~\ref{JKcmd}
shows the approximate limit of the combined $JK/Spitzer$ catalog. It
corresponds to $\sim 0.01 M_{\odot}$ at $A_V=0$ and $\sim 0.03 M_{\odot}$ at $A_V=30$, based on COND03 \citep{baraffe03}
and DUSTY00 \citep{chabrier00} evolutionary models.

Based on the ISOCAM survey of the main $\rho$ Oph core, \citet{bontemps01} find that the fraction of objects with disks
among the embedded cluster YSO population is $\sim 54\%$. From a small sample of 9 BDs, \citet{jayawardhana03} calculate the disk fraction among the BD population of $\sim (67 \pm 27)\%$. The number of the missing BDs with disks in our sample is $\sim 4$. Accounting for the disk fraction, we estimate that the total number of missing BDs $\sim 7$.

\subsection{Substellar population in $\rho$ Oph}
\label{missing}
Based on the results of our survey, and combined with literature data, we can try to put
limits on the total number of substellar objects in $\rho$ Oph. 
However, we must note that $\rho$ Oph, with its high and strongly
varying extinction, is a difficult field to search for BDs.
All current surveys, including SONYC, have to be considered incomplete. 

SONYC was designed to mitigate selection biases that are typically induced by
commonly used mid-infrared-based surveys (sources with discs), or narrow-band methane surveys (T-dwarfs). Our primary
means of selecting candidates for spectroscopy is broad-band
optical and near-infrared imaging that aims to detect the photosphere. This selection
is clearly not effective in the areas with high extinction, and is also significantly
affected by the background contamination.
However, in an attempt to get an estimate of the number of missing substellar objects in our survey area,
without being biased towards the disk objects, we base the following discussion 
only on the optical-NIR candidate selection. 

Figure~\ref{IJcmd} shows the ($i'$, $i'-J$) color-magnitude diagram for all the sources from our $i'JK$ catalogs. Black circles show 309 candidates from the MOIRCS $JK$ selection, while purple open circles show sources with $JK$ photometry from 2MASS. Crosses show SONYC spectroscopic targets, with confirmed VLMOs shown as squares.
Among the SONYC VLMOs, the object found at $(i'$,$i' - J)=(24.1, 5.2)$ is identified as a low-mass star, while the remaining five are BDs.
 We also show spectroscopically confirmed stellar members (stars) and BDs (diamonds) from the literature. Distribution of the objects with spectra and confirmed substellar member candidates shows that the criterion used for the candidate selection is fairly conservative.

Candidates from the optical-NIR selection that have been spectroscopically followed up
come from the $i'JK_S$ catalog based
on the MOIRCS $JK$ photometry. This candidate list contains 309 objects. We have obtained spectra for 102 of these
objects. Five were confirmed as likely BDs. 
Thus, the spectroscopic
follow-up suggests that our catalog could contain $(309-102)\times 5/102 \approx 10$ additional BDs. This number might actually be smaller because, as evident from Figure~\ref{IJcmd}, the confirmed BDs tend to be located
to the right of the DUSTY00 1 Myr isochrone. Based on this observation it might be reasonable to shift the selection cutoff line slightly towards the right (green dash-dotted line). The probability of finding any new members in the strip between our old selection line and the new one is expected to be very low. The new selection cutoff was chosen to lie approximately at the same distance from the two isochrones shown in Figure~\ref{IJcmd}. The new candidate list then contains 151 objects, 75 of which have spectra. The expected number of missing BDs is then $(151-75)\times 5/75 \approx 5$.  

The combined $i'JK$ catalog is shallower than the $i'$-band catalog and its completeness limit is at $i'\approx 23.4$.  
The majority of the optically selected candidates lie in the area with low extinction (see Figure~\ref{spatial}): 89\% of the candidates are located at $A_V \leq 15$, as judged from the COMPLETE map. However, as shown in Figure~5 of \citet{geers11}, all objects from the MOIRCS $i'JK$ catalog, with the exception of only two sources, probably suffer optical extinctions of less than 15. The extinction for each source was calculated based on the observed $(J-K)$ colors, assuming an intrinsic $(J-K)$ of 1.
Thus, even in the regions with apparently large extinction, our optical-NIR observations seem to be penetrating the cloud only down to a certain depth, corresponding to $A_V$ of about 15. The completeness limit of $i'\approx 23.4$ corresponds to masses $0.003 - 0.03 M_{\odot}$ for $A_V = 0 - 15$, according to the DUSTY00 and COND03 models for 1 Myr old objects at the distance of $\rho$ Oph. We conclude that the number of missing objects in our survey area down to this mass and extinction limit is 5-10. 
This number should be added to the number of missing objects estimated from the $JK-Spitzer$ data in Section~\ref{S_MIR}. The analysis presented
in this section is based on the 309-candidate MOIRCS $i'JK$ catalog and does not include the 2MASS $i'JK$ selection. The $JK-Spitzer$ candidate selection
is, on the other hand, almost entirely based on the 2MASS JK photometry (72/83 candidates). In Figure~\ref{JKcmd}, the $JK-Spitzer$ candidates are 
mostly found above the grey line, while $\sim 85 \%$ of the MOIRCS $i'JK$ candidates lie below it. The  two catalogs overlap for only 3 objects.

In summary, we conclude that the number of missing substellar objects in our survey area is 12-17.
The actual number of missing substellar objects is probably higher, as we certainly might be missing objects at lower masses and higher extinctions.

\subsection{Ratio of stars to brown dwarfs}
Based on the currently available data we can put constraints on the ratio of the number of low-mass
stars to the number of BDs in our survey area. 
\citet{andersen08} use a mass range $m = (0.08 - 1) M_{\odot}$ for stars and $m = (0.03 - 0.08) M_{\odot}$ for BDs; we call this ratio $R_1$. Other authors use $m = (0.08 - 10) M_{\odot}$ for stars and $m = (0.02 - 0.08) M_{\odot}$ for BDs, hereafter called $R_2$ 
\citep{briceno02, muench02, luhman03}. 

The two most complete estimates of the stellar population in $\rho$ Oph are those from \citet{bontemps01} and \citet{wilking08}. \citet{bontemps01} report 123 Class II YSOs, 95 of which are located within the borders of our optical survey. From their Figure~8, we infer $\sim$80 YSOs in the mass range $(0.08 - 1) M_{\odot}$, and $\sim$84 in the mass range $(0.08 - 10) M_{\odot}$. We scale these numbers with the fraction of all sources located in our FOV (95/123), and correct for the binary fraction of about $29\%$ \citep{ratzka05}. The number of stars in our FOV is then $\sim 80$ and $\sim 84$ for mass ranges $(0.08 - 1) M_{\odot}$ and $(0.08 - 10) M_{\odot}$, respectively.
From the member list in \citet{wilking08} we discard objects with spectral type M6 or later, restrict the coordinates to match the area covered by our survey, add the three VLMOs discovered in the framework of SONYC, and four stellar-mass members recently identified by \citet{erickson11}.  
We count 86 low-mass stars with spectral type estimates between G2 and M6 (i.e. with masses between 0.08 and 1 $M_{\odot}$, and 90 stars in the mass range $(0.08 - 10) M_{\odot}$.

There are currently 26 spectroscopically confirmed probable BDs within the borders of our survey (20 previously known and 6 found in this work). 
We calculated absolute J-band magnitudes for the sample of BDs using the $A_V$ estimated from $J-K$ colors, reddening law from \citet{cardelli89}, and by assuming the distance of 125~pc. We then compare the results with the COND03 and DUSTY00 1~Myr isochrones. There are 21 objects with $m \gtrsim 0.03 M_{\odot}$, and 23 with $m \gtrsim 0.02 M_{\odot}$. Among these there are 8 objects for which we obtain mass estimates above $0.08 M_{\odot}$; 7 come from the list of \citet{wilking08} and one is our object SONYC-RhoOph-5. All of these objects have spectral type around M6, i.e. they lie around the boundary separating low-mass stars from BDs. Considering that the mass estimates are subject to a number of uncertainties (arising from the uncertainties in cluster distance, age, extinction, photometry, effective temperature, and the models themselves), all of these 8 objects could in fact be either low-mass stars or BDs. 
Thus, for the calculation of the ratios $R_1$ and $R_2$, we assume the following numbers:\\ 
\\
(1) stars with $m=(0.08 - 1) M_{\odot}$: $80 - 94$, \\
(2) stars with $m=(0.08 - 10) M_{\odot}$: $84 - 98$, \\
(3) BDs with $m=(0.03 - 0.08) M_{\odot}$: $13 - 21$,\\
(4) BDs with $m=(0.02 - 0.08) M_{\odot}$: $15 - 23$.\\
\\
To calculate the ratios $R_1$ and $R_2$ we adopt the mean values of the intervals specified above. A reasonable indication of statistical uncertainties associated with these numbers is provided by Poisson statistics, i.e. we assume the standard deviation $\sigma = \sqrt\mu$, where $\mu$ stands for the mean of each interval.

Finally, we arrive at $R_1=5.1 \pm 1.4$ and $R_2 = 4.8 \pm 1.2$. 
This result is in line with the values from the literature (3.3 - 8.5; \citealt{andersen08}), although higher than the value observed in NGC1333 ($R_1=R_2=2.3$; \citealt{scholz11}), and in the ONC ($R_1=2.4$; \citealt{andersen11}). 
Note that the values quoted here and in \citet{andersen08} should be treated as upper limits. The total number of BDs is certainly higher than the values given above due to the lower mass cutoff set at $0.02-0.03 M_{\odot}$ and the missing BDs.

\subsection{Spatial distribution and disks}
\begin{figure*}
\center
\includegraphics[width=15cm,angle=0]{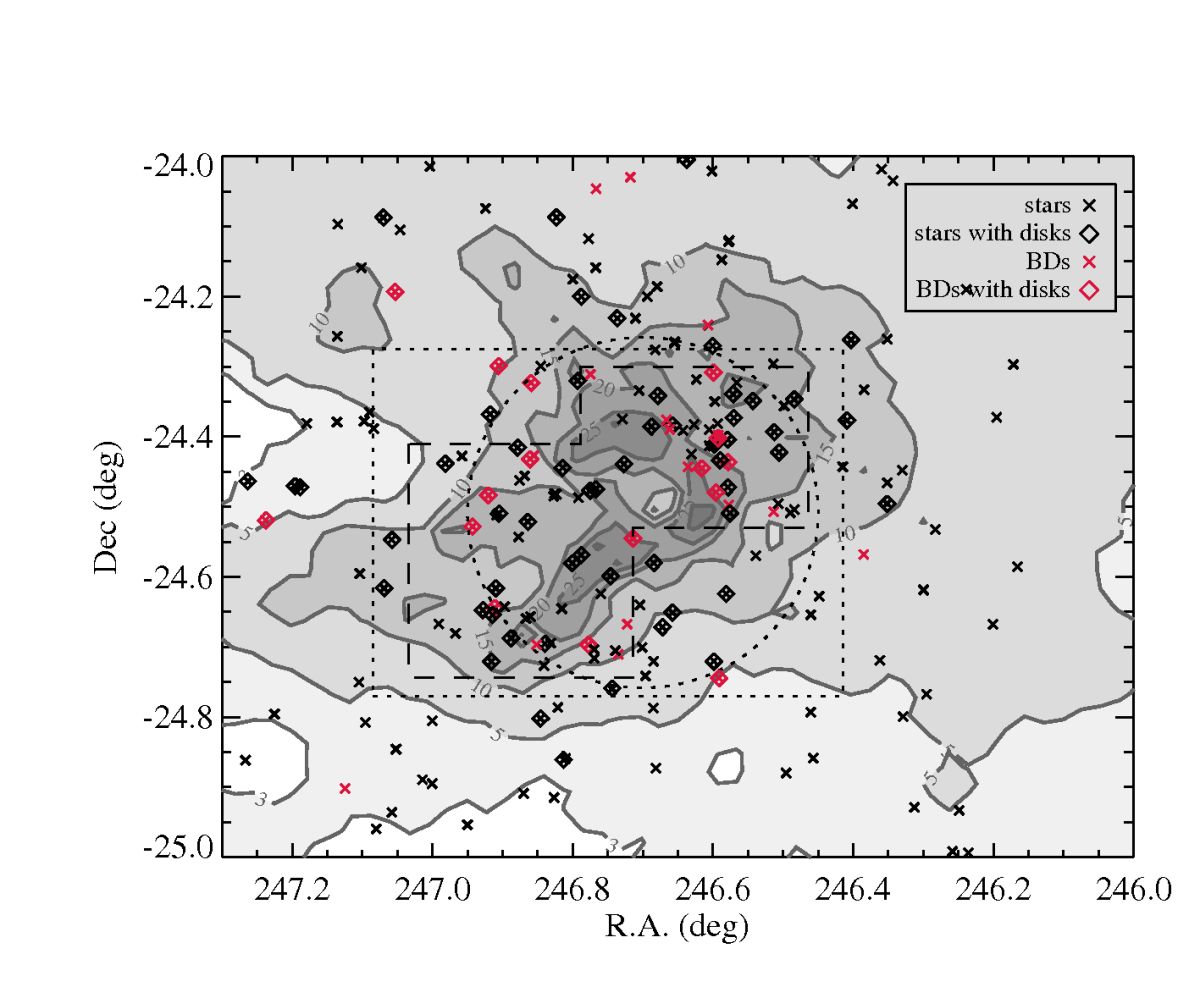}
\caption{Spatial distribution of spectroscopically confirmed members in $\rho$ Oph. 
Contours and lines are same as in Figure~\ref{spatial}, with addition of the lowest extinction contour. 
Stellar members and BD are shown as black and red symbols, respectively. Crosses mark objects without
disks, and diamonds those with disks (i.e. showing MIR excess).}
\label{spatial_disks}
\end{figure*}

In Figure~\ref{spatial_disks} we show the projected distribution of spectroscopically confirmed
stellar members and BDs in $\rho$ Oph. Stellar-mass members (black crosses) are from \citet{wilking08} and \citet{erickson11}, with three
additional objects identified as very-low-mass stars in the framework of SONYC. Red crosses show spectroscopically confirmed BDs from \citet{wilking08}, \citet{alvesdeoliveira10}, \citet{marsh10}, \citet{geers11}, \citet{erickson11} and this work. Both catalogs of stellar and substellar-mass members were cross-matched with the catalog containing objects showing MIR color-excess, which is interpreted as the evidence for an existence of a circumstellar disk around these objects. 
The stellar population clearly shows clustering towards the inner part of the L1688 core. There are 80 objects within the boundaries of our optical survey (dotted rectangle), and 65 objects outside of it (in the area roughly twice as large).
At a first glance it might seem that even stronger clustering exists among the BD population. However, this can be attributed to the lack of a uniform BD survey in the entire cluster region. Most of the existing BD surveys, including this one, have been concentrated on the inner core of the cloud, and only two surveys actually probed the larger area \citep{cushing00, wilking05}. Therefore, the larger star-to-BD ratio in the outskirts of the cluster is probably not real. 
Within our FOV, there is apparently no significant difference between the spatial distribution of stellar and substellar populations.   

A comparison between the two populations can also be made based on the presence of disk excess. Here we again concentrate on the area within the $i'$-band survey boundaries. The ratio of the number of stellar members with disks, to the total number of stellar members is $37/90=(41 \pm 8)\%$. This number probably represents the lower limit for the disk fraction, because of the relatively rigorous criteria on photometry that were applied to the 2MASS data. For the $2MASS-Spitzer$ source matching, we request photometric quality flags for $JK$ to be at least ``B", i .e. we select only detections with $SNR>7$ and photometric uncertainties below $0.1551$ mag. Relaxing this criterion by allowing selection of the flags ``C'' and ``D'', we end up with the stellar disk fraction of $41/90=(46 \pm 9)\%$. Among the BD population in our FOV, we find that the fraction of sources exhibiting MIR excess is 
$15/26=(58 \pm 19)\%$, in agreement with results from \citet{jayawardhana03}. The YSO sample from \citet{bontemps01} includes both stars and BDs, down to $\sim0.055M_{\odot}$. Combining our numbers for stars and BDs, we get the disk fraction of $(45 \pm 7)\%$, in agreement with \citet{bontemps01}.

Similar results were obtained in other star forming regions. For $\sigma$ Ori, Chamaeleon-I, and IC~348, disk fractions are found to be $\sim 40\%$ and $\sim 50 - 60 \%$ for low-mass stars and brown dwarfs, respectively \citep{luhman08b}. In NGC~1333, \citet{scholz11} find an upper limit on the BD disk fraction of $66\%$. $\sigma$ Ori, Cha-I, and IC~348  
are somewhat older than $\rho$ Oph and NGC~1333 (2-3 Myr vs $\lesssim$ 1 Myr), indicating that there is no significant change in the disk fractions over the first 3 Myr of the cluster lifetime.

\section{Conclusions}
\label{concl}
In this work we have built upon the large-scale optical and near-infrared survey of the L1688
core in $\rho$ Ophiuchi presented in \citet{geers11}. From the optical+NIR photometry, 309
objects were selected as candidate substellar cluster members. 58/309 of these objects
were targeted for spectroscopy, and one was confirmed as a substellar mass object \citep{geers11}. 
Here we report the second round of spectroscopic follow-up, where we obtained spectra of additional 46
candidates from the same list. Among these, we find 4 candidate substellar 
objects with $T_{\mathrm{eff}} \leq 3000$K, 3 of which are for the first time reported here (SONYC-RhoOph-3, 6, 7).

From MOIRCS, 2MASS and $Spitzer$ photometry, \citet{geers11} established an additional sample of 26 candidate substellar members with disk excess.
12 of these appear as spectroscopically confirmed BDs with spectral type later than M5 in the literature.
We obtained spectra of 17 $Spitzer$ BD candidates, including 9 previously confirmed BDs. We report 5 newly discovered 
very-low-mass probable members of $\rho$ Oph, including 3 BDs (SONYC-RhoOph-5, 8, 10). From the list of the previously confirmed objects, we re-confirm 7 out of 9 with $T_{\mathrm{eff}} \leq 3200$K. 

In summary, we report the discovery of 8 new very-low-mass candidate members in $\rho$ Oph with
$T_{\mathrm{eff}}$ between 2550 and 3200$\,$K. Six of these objects are classified as brown dwarfs, and
comprise one fifth of the entire currently known substellar population in $\rho$ Oph. 
While it might be premature to derive the substellar IMF in this cluster, as the high and variable
extinction seriously hampers most attempts to detect objects at the lowest masses, we estimate that the upper limit
on the ratio of the number of low-mass stars to the number of BDs is
in line with other nearby star forming regions.
By analysis of the MIR excess among the spectroscopically confirmed members of $\rho$~Oph, we 
find the disk fractions to be $\sim 40 \%$ and $\sim 60\%$ for stars and BDs, respectively. This is again in line with the results from the other nearby star forming regions.

\acknowledgments
The authors would like to thank the Subaru staff, especially 
Dr. Naoyuki Tamura and Dr. Kentaro Aoki, for the assistance during
the observations and their preparation. We are grateful to Ms. Yuuki Moritani, 
Mr. Kiyoto Yabe and Prof. Fumihide Iwamuro (Kyoto University) for their help
with the FMOS data reduction.
The research was supported in part by grants from
the Natural Sciences and Engineering Research Council
(NSERC) of Canada to RJ.
RJ also acknowledges support from a Royal Netherlands Academy of Arts
and Sciences (KNAW) visiting professorship.
This work was also supported in part by the Science Foundation Ireland  
within the Research Frontiers Programme under grant no. 10/RFP/AST2780.
MT is supported by MEXT Grant-in-Aid for Specially Promoted Research. 
This publication makes use of data products from the Two Micron All
Sky Survey, which is a joint project of the University of Massachusetts
and the Infrared Processing and Analysis Center/California Institute of
Technology, funded by the National Aeronautics and Space Administration
and the National Science Foundation. This research has benefitted from 
the SpeX Prism Spectral Libraries, maintained by Adam Burgasser 
at 
{\tt http://pono.ucsd.edu/\textasciitilde adam/browndwarfs/spexprism/}.


\clearpage

\end{document}